\documentclass[a4paper, amsfonts, amssymb, amsmath, reprint, showkeys, twoside, prl,superscriptaddress,hidelinks]{revtex4-2}
\usepackage[english]{babel}
\usepackage[utf8]{inputenc}
\usepackage{graphicx}
\usepackage{bbm}
\usepackage{appendix}
\usepackage{amsthm}
\usepackage{mathtools}
\usepackage{physics}
\usepackage{xcolor}
\usepackage{graphicx}
\usepackage{adjustbox}
\usepackage{placeins}
\usepackage[T1]{fontenc}
\usepackage{lipsum}
\usepackage{csquotes}
\usepackage{color}
\usepackage{braket}
\usepackage{amsmath}
\usepackage{float}
\usepackage{hyperref}
\usepackage[caption=false]{subfig}
\newcommand{\be}{\begin{equation}}
\newcommand{\ee}{\end{equation}}

\begin{document}

\title{
Accelerating cosmology from a holographic wormhole
}

\author{Stefano Antonini}
\email{santonin@umd.edu}
\affiliation{Maryland Center for Fundamental Physics, University of Maryland, College Park, MD 20742, USA}
\author{Petar Simidzija}
\email{psimidzija@phas.ubc.ca}
\affiliation{Department of Physics and Astronomy, University of British Columbia, Vancouver, BC V6T 1Z1, Canada.}
\author{Brian Swingle}
\email{bswingle@brandeis.edu}
\affiliation{Maryland Center for Fundamental Physics, University of Maryland, College Park, MD 20742, USA}
\affiliation{Brandeis University, Waltham, MA 02453, USA}
\author{Mark Van Raamsdonk}
\email{mav@phas.ubc.ca}
\affiliation{Department of Physics and Astronomy, University of British Columbia, Vancouver, BC V6T 1Z1, Canada.}

\begin{abstract}
    We consider cosmological models in which the cosmology is related via analytic continuation to a Euclidean asymptotically AdS planar wormhole geometry defined holographically via a pair of three-dimensional Euclidean CFTs. We argue that these models can generically give rise to an accelerating phase for the cosmology due to the potential energy of scalar fields associated with relevant scalar operators in the CFT. We explain how cosmological observables are related to observables in the wormhole spacetime and argue that this leads to a novel perspective on naturalness puzzles in cosmology.
\end{abstract}

\maketitle

\paragraph{Introduction.}

It is a major open question in theoretical physics to come up with a complete quantum gravity model of four-dimensional big bang cosmology. Current cosmological models rely on effective field theory which does not include quantum gravity effects and which cannot provide insight into the big bang or the structure of the effective theory. 
Via the AdS/CFT correspondence (or "holography") we do have fully microscopic models of quantum gravity for simpler spacetimes which are asymptotically empty and negatively curved. It is natural to ask whether such models can also describe cosmological spacetimes.

The effective gravitational theories that arise in holographic models generally have negative cosmological constant. Cosmological solutions with matter and/or radiation and a negative cosmological constant are time-symmetic big bang / big crunch spacetimes. A hint for how holography might describe these is that the analytic continuation of the cosmologies to Euclidean proper time $\tau$ give Euclidean wormholes with asymptotically AdS regions at $\tau = \pm \infty$ \cite{Maldacena:2004rf}. In \cite{Cooper2018, Antonini2019,Antonini:2021xar,VanRaamsdonk:2020tlr, VanRaamsdonk:2021qgv,Antonini2022}, we proposed to holographically describe such Euclidean wormholes via a pair of 3D Euclidean holographic CFTs coupled by a higher dimensional field theory, and to use this Euclidean construction to define a state for the Lorentzian cosmology (similar to the construction of Hartle and Hawking).

The negative cosmological constant present in the effective theories that arise in this construction might appear to be in tension with the observations that our universe is accelerating \cite{perlmutter1999measurements, riess1998observational}. However, in this paper, we argue that holographic wormhole models of cosmology almost always have time-dependent scalar fields, and that the potential energy from these scalar fields can generically give rise to an accelerating phase for the cosmology before the recollapse. Thus, holographic wormhole models have the potential not only to give complete quantum gravity models of cosmology, but also to give models that are realistic!

We also point out some intriguing implications of this framework for cosmological observables. The cosmology has a preferred quantum state determined by the choice of effective field theory alone. This state exhibits time-reversal symmetry and observables that are related by analytic continuation to vacuum observables in a static, weakly-curved asymptotically AdS spacetime. Exploiting this relation, the computation of cosmological observables can be carried out without any detailed knowledge of the UV completion or of the physics near the big bang, though the underlying string theory description provides useful constraints on the effective theory. The framework potentially resolves various naturalness puzzles without the need for inflation. It also provides candidate explanations for dark energy and its smallness, the cosmological coincidence problem, and the emergence of the standard model from an underlying supersymmetric theory.

More details can be found in the companion papers \cite{Antonini2022,Antonini:2022fna}. Other approaches to cosmology that make use of some features of our framework include \cite{McFadden:2009fg,hartle2012accelerated,boyle2018c,Penington:2019kki, Dong:2020uxp, Chen:2020tes,Boyle:2018rgh}, as described in  \cite{Antonini2022}.

\paragraph{Time-symmetric cosmology from Euclidean wormholes}

We begin by considering flat cosmological solutions with negative cosmological constant, matter, and radiation. In the metric
\be
\label{Cbackground}
ds^2 = -dt^2 + a^2(t) d\vec{x}^2 \; ,
\ee
the scale factor $a(t)$ initially increases after the big bang, but as radiation and matter dilute, the negative cosmological constant becomes increasingly important, resulting in deceleration and eventual collapse. The full solutions are time-reversal symmetric, and can be analytically continued to real reflection-symmetric Euclidean spacetimes
\be
\label{Ebackground}
ds^2 = d\tau^2 + a_E^2(\tau) d\vec{x}^2 \; .
\ee
where $a_E(\tau) = a(i \tau)$. The Euclidean scale factor $a_E(\tau)$ increases away from this point, and has asymptotically AdS behavior for $\tau \to \pm \infty$. The metric (\ref{Ebackground}) therefore describes a Euclidean AdS wormhole connecting two distinct asymptotic boundaries.

In quantum field theory and quantum gravity, the Euclidean version of a theory is often employed in order to construct a natural state of the Lorentzian theory. For example, the path integral for a Euclidean quantum field theory on $\mathbb{R}^n$ constructs the vacuum state of the corresponding Lorentzian theory on $\mathbb{R}^{n-1,1}$. Similarly, we propose to use the Euclidean gravity path integral with asymptotically AdS boundary conditions in the past and future to define a special state for the cosmology. This is similar to the Hartle-Hawking proposal for the wavefunction of the universe \cite{Hartle:1983ai}, but we have asymptotically AdS boundary conditions in the Euclidean past instead of the no-boundary condition of Hartle and Hawking. To fully define this gravitational path integral, we need to provide some UV completion. The asymptotically AdS behavior suggests that this can be accomplished via holography, making use of an underlying three-dimensional CFT associated with the asymptotically AdS boundary \cite{Maldacena:1997re,Witten1998,Aharony1999}. For the full geometry with two asymptotically AdS boundaries, the proposal is to make use of a pair of 3D Euclidean CFTs coupled together by an auxiliary higher-dimensional field theory. Such holographic models are discussed in detail in \cite{VanRaamsdonk:2021qgv,Antonini2022}, however, the only input we will need from this underlying description are the four-dimensional effective gravitational theory that it provides and the boundary physics of the fields in the effective theory at the asymptotically AdS boundaries.

\begin{figure}
    \centering
    \includegraphics[width=\columnwidth]{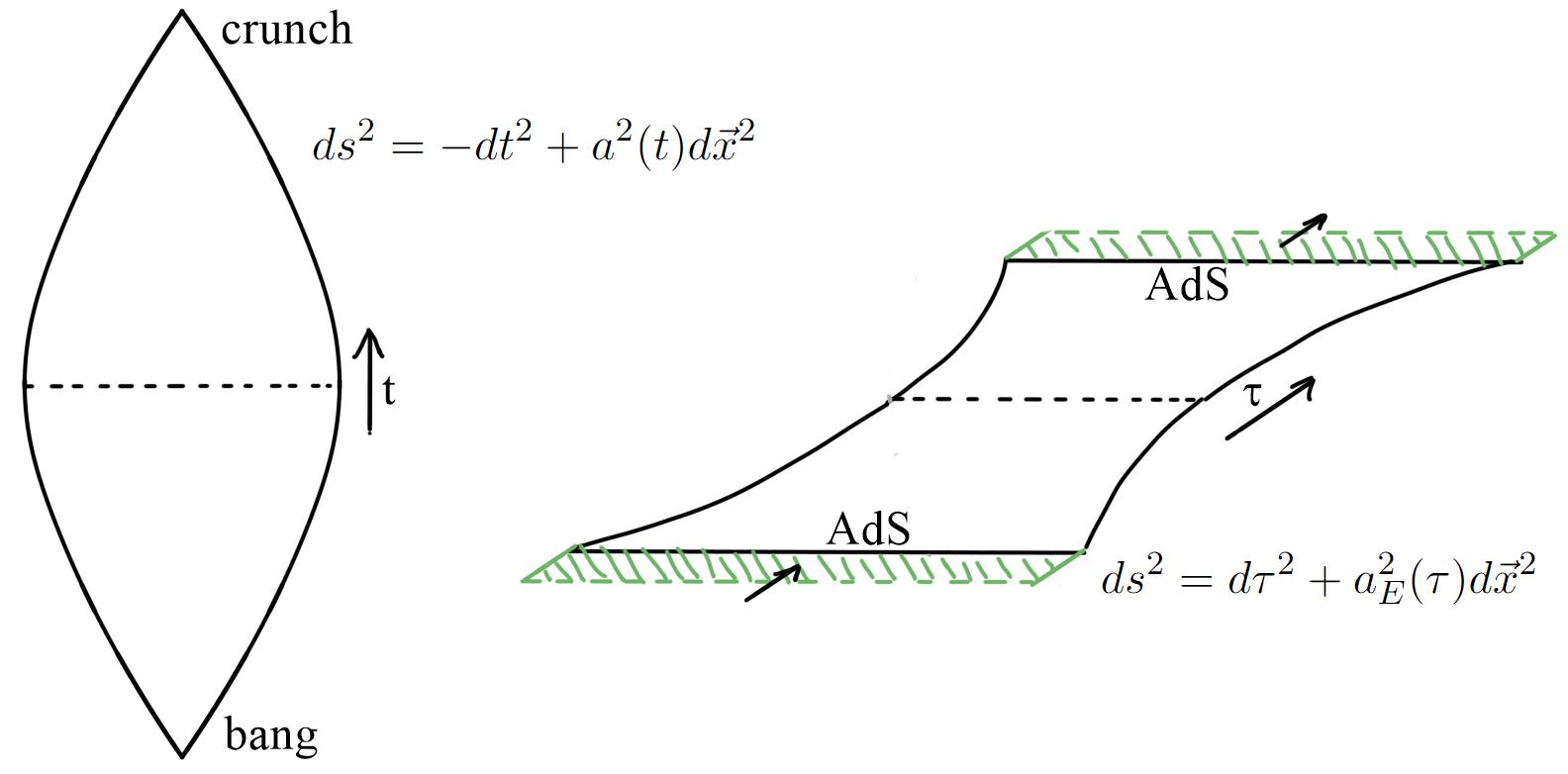}
    \caption{Left: time-symmetric $\Lambda < 0$ cosmology. Right: replacing $t = i \tau$ gives a Euclidean wormhole with two asymptotically AdS regions. The effective field theory description of such a wormhole may require coupling the fields in the two asymptotic regions via an auxiliary non-gravitational field theory (green).}
    \label{Shortfig}
\end{figure}

To summarize, we
\begin{itemize}
\item
Start with a 4D gravitational effective field theory. The choice here is constrained by requiring some consistent dual CFT. In particular, the requirement of asymptotically AdS solutions requires that the scalar potential should have an extremum with a negative value. 
\item
Specify some boundary conditions for the fields in the Euclidean version of this theory at a pair of asymptotically AdS boundaries. This is a crucial step that we discuss further below.
\item
Find a reflection-symmetric solution for the background geometry that connects the two asymptotically AdS regions. This gives the Euclidean background geometry (\ref{Ebackground}).
\item
Analytically continue this to find the Lorentzian background cosmology (\ref{Cbackground}). Or, equivalently, use the metric and fields on the reflection-invariant slice to define initial data for the background fields in the cosmology.
\item
Use the effective field theory on the Euclidean background to define a state of the quantum fields in the cosmology.
\end{itemize}

\paragraph{Time dependent scalars are generic}

Remarkably, time-varying scalars with a very flat potential are generic in this holographic setup. Effective gravitational theories dual to holographic 3D conformal theories typically have some scalars with negative mass squared; these are in one-to-one correspondence with relevant scalar operators in the dual CFT. In generic asymptotically AdS solutions to the effective theory, such negative mass scalars develop expectation values as we move away from the asymptotically AdS boundary \cite{SUGRA4,Karndumri:2022rlf}.

The action for gravitational and scalar fields can be written as
\be
\label{scalaraction}
S = \frac{1}{2k} \int d^4x \sqrt{g}(R -  g^{ab}\partial_a \phi \partial_b \phi - \frac{6}{L^2} V(\phi)) + S_{matter}
\ee
where $k=8\pi G$. Here, the scalar field and its potential $V(\phi)$ are dimensionless, and the cosmological constant $\Lambda = -3/L^2$ has been included in the scalar potential, which we can write as 
\be
\label{eq:Vpot}
V(\phi) =   -1 + \frac{1}{2} \hat{m}^2 \phi^2 + V_{int}(\phi) \; .
\ee
For effective actions dual to CFTs, it is natural that the parameters in this potential (including $\hat{m}$) are dimensionless numbers of order one. In particular, we have that
\be
 -\frac{9}{4} \le \hat{m}^2 < 0
\ee
for scalars associated with relevant CFT operators (the lower bound also follows from requiring stability of asymptotically AdS gravitational solutions). Since the Planck scale appears only in front of the action, the natural length/time scales for variations of the scalar fields in the geometry is the same as the cosmological constant scale.
The cosmological constant scale sets the age of the universe in the cosmology picture, so the conclusion is that the time scale for variation of dark energy produced by these scalar fields is the same as the age of the universe.

\begin{figure}
  \centering
  \includegraphics[width=0.8 \columnwidth]{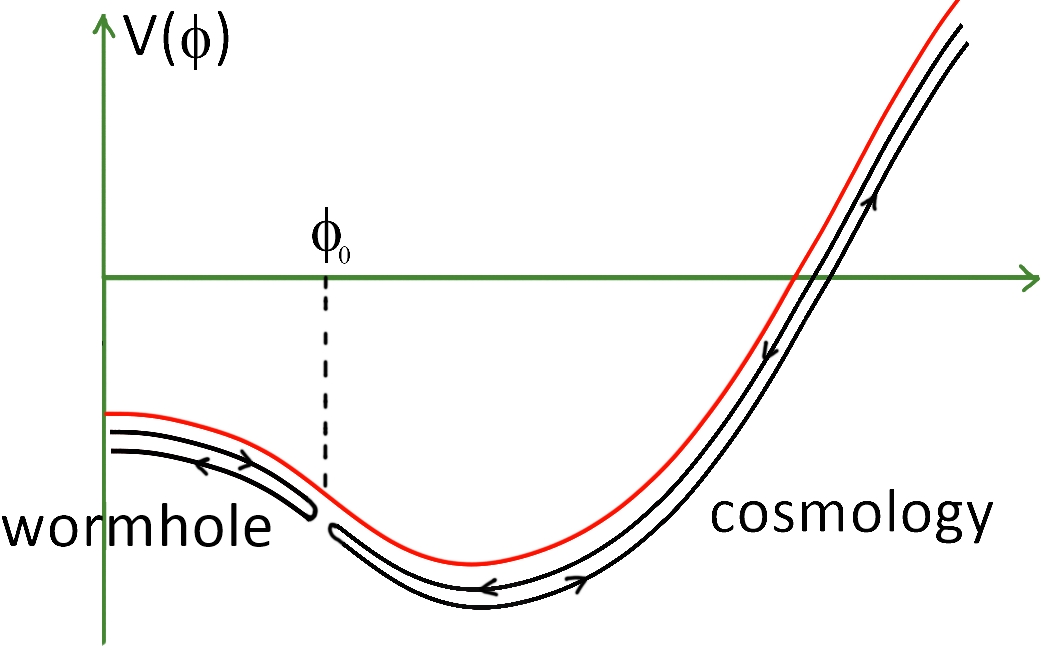}
  \caption{Typical potential for scalar field associated relevant operator in the dual CFT. Paths indicate scalar evolution in the wormhole from $\tau = -\infty$ to $\tau = \infty$ and in the cosmology from big bang to big crunch. The initial positive values of the potential can give rise to a phase of accelerated expansion before deceleration and collapse.}
\label{potentialfig}
\end{figure}

\paragraph{Accelerating cosmology from scalars.}

The scalar field expectation values vary with radial position in the wormhole solutions, and with time in the cosmology. The potential energy from these scalars has the interpretation of a time-dependent dark energy. A typical potential in one of these negative mass squared directions is shown in Figure \ref{potentialfig}; the negative curvature at $\phi = 0$ corresponds to the negative mass squared of the scalar, while the increase for larger $\phi$ comes from interaction terms in the potential that increase for large $\phi$ in a stable theory. The evolution of the scalar from the middle of the wormhole to the asymptotically AdS boundary is the same as a damped motion of a particle with coordinate $\phi$ in a potential $-V(\phi)$, so typically descends from some value $\phi_0$ with negative $V$ to the asymptotic value $\phi = 0$ with a smaller magnitude negative $V$. The evolution of the scalar in the expanding phase of the cosmology is the same as a damped motion of a particle with coordinate $\phi$ in a potential $V(\phi)$. Thus starting from the early universe, the scalar potential may decrease from positive values before becoming negative and reaching $\phi_0$ at the time when $\dot{a}=0$. 

We will now see that these positive values of the scalar field can generically (i.e. without fine tuning) give rise to a phase of accelerated expansion before the collapse; in a realistic example, this could be the present accelerating phase of the universe. We take $\hat{m}^2 = -9/4$ and consider a one-parameter family of interaction terms
\be\label{eq:intpot}
V_{\text{int}}(\phi) = e^{g \phi^2} - g \phi^2 - 1
\ee
where the last two terms are subtracted off to give an interaction potential starting at order $\phi^4$. 

\begin{figure}
    \centering
    \includegraphics[width=\columnwidth]{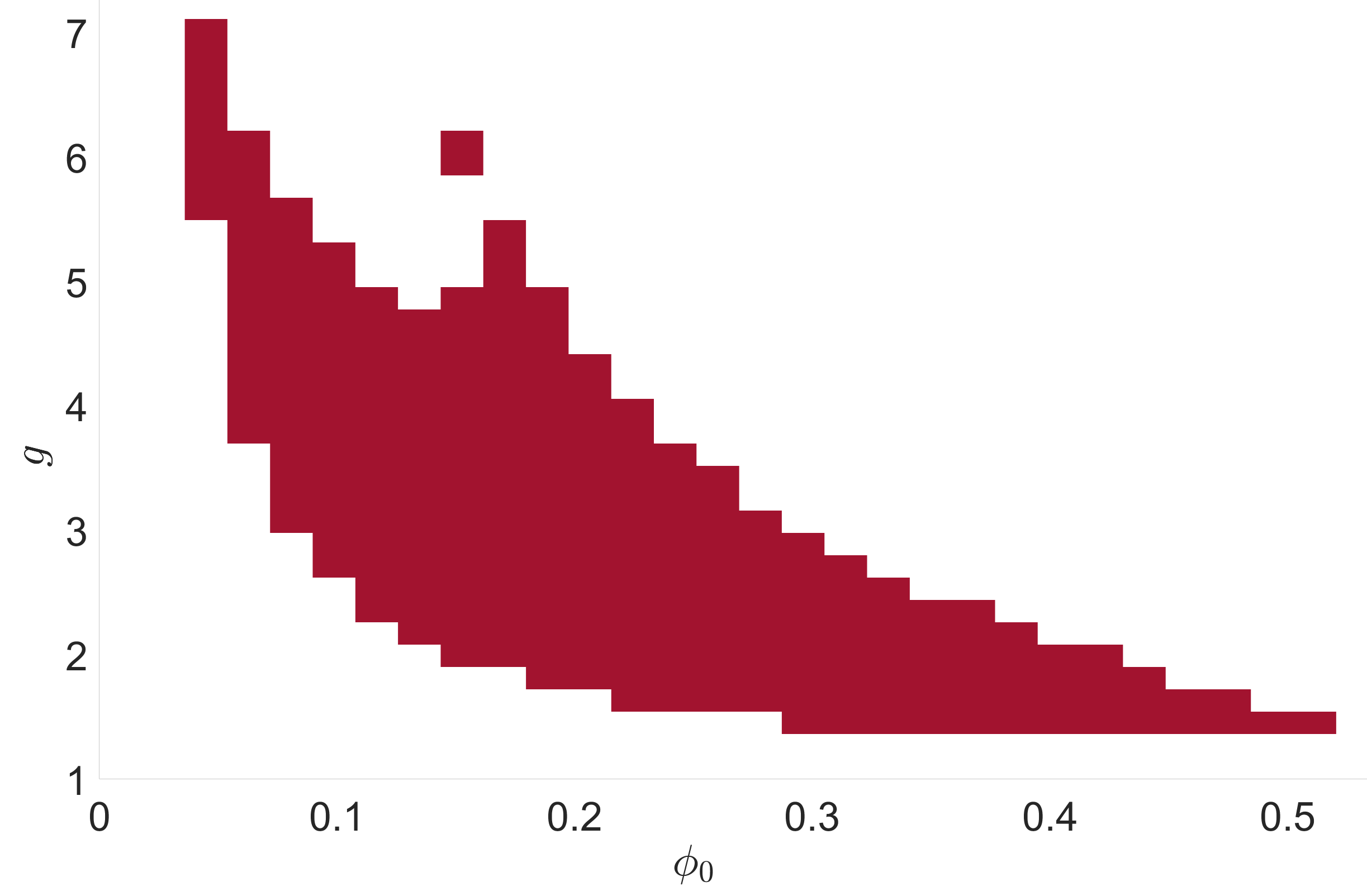}
    \caption{Space of model parameter $g$ and initial condition $\phi_0$ for the exponential potential of (\ref{eq:Vpot}) and (\ref{eq:intpot}).}
    \label{fig:paramspace}
\end{figure}

Numerically integrating to find the solutions of the coupled Friedmann equation and scalar equations for a given coupling parameter $g$ and initial condition $\phi_0$, we find that a significant region of the parameter space includes a period of accelerated expansion before the recollapse (Figure \ref{fig:paramspace}).

Thus, cosmological acceleration can be obtained without significant fine-tuning of the model or initial conditions. More details, as well as examples of potentials giving rise to scale factors in agreement with observations and admitting a well-defined analytic continuation to a Euclidean wormhole, can be found in \cite{Antonini:2022fna}.

\paragraph{Extracting cosmological observables.}

By our construction, the background cosmological spacetime and the correlation functions in the cosmology (which encode structure, the CMB, etc...) are related by analytic continuation of the time coordinate $t$ to observables in the Euclidean theory. Via a simpler analytic continuation (replacing $x \to ix$ for one of the translation-invariant $\vec{x}$ coordinates), they are also related to the vacuum physics of the Lorentzian theory on a third spacetime, 
\be
\label{LW}
ds^2 = d\tau^2 + a_E^2(\tau) d x_\mu d x^\mu \; ,
\ee
which is a static spacetime with a pair of asymptotically AdS regions at $\tau = \pm \infty$. We will refer to this as the Lorentzian wormhole geometry since it has two asymptotically AdS regions connected through the interior \cite{Antonini:2022opp}.

In practice, the simplest way to extract predictions for the cosmological physics is to compute observables in either the Euclidean or Lorentzian asymptotically AdS geometries and then analytically continue these back to the cosmological spacetime. In these other descriptions, the background geometries are weakly curved everywhere, so it is not necessary to know about the details of physics near the big bang (including the details of the UV completion of gravity) to compute observables. Further, in the Lorentzian wormhole picture, we only need to understand observables in the vacuum state of the theory; remarkably, these should contain everything we wish to know about the cosmological physics. Finally, the presence in the wormhole geometries of the two asymptotic AdS boundaries allows us to relate cosmological observables to observables in the dual UV-complete microscopic theory using the standard holographic dictionary.

As an example, equal time correlation functions $\langle T_{tt}(x_1,t) \cdots  T_{tt}(x_n,t) \rangle$ of the energy density in the cosmology are obtained by analytic continuation $\tau \to it$ from vacuum correlation functions $\langle T_{\tau \tau }(x_1,\tau) \cdots  T_{\tau \tau}(x_n,\tau) \rangle$  of the stress tensor component $T_{\tau \tau}$ for the same effective field theory in the geometry (\ref{LW}), where operators are chosen to lie in the 2D plane that remains spatial in all three pictures. The cosmological correlators obtained in this way correspond to quantum correlators in the full wavefunction. These can be understood as an average over the ensemble of possible classical cosmologies described by the wavefunction. We can equivalently understand the correlators as telling us about a spatial average of correlators in a typical cosmology in the ensemble (see e.g. \cite{weinberg2008cosmology}); in this way we can try to connect with observations in our own universe. %

\paragraph{The effective field theory.}

If this framework can describe realistic cosmology, the underlying gravitational effective field theory should be some extension of the Standard Model. The existence of a holographic description suggests that this extension should be some 4D supergravity theory with a gauge group G that will ultimately be broken to give $SU(3) \times SU(2) \times U(1)$. The asymptotically AdS regions correspond to a solution of this effective theory where supersymmetry and gauge symmetry are preserved and the scalar fields lie at some extremum of the potential with negative value; away from the boundaries (and in the cosmological solution) the scalars have varying expectation values that contribute to gauge and supersymmetry breaking as described below.

\paragraph{The cosmological constant.}

The extremum of the scalar potential associated with the AdS boundaries sets the fundamental cosmological constant for the theory. This is directly related to the number of degrees of freedom in the dual CFT. Thus, the smallness of the cosmological constant (and the largeness of the universe relative to the Planck scale) is ultimately related to having an underlying CFT with a very large number of degrees of freedom. As discussed further in \cite{Antonini2022}, it is likely important that the underlying CFT is chosen to be supersymmetric so that the bulk cosmological constant does not receive large quantum corrections. In this case, the underlying bulk effective theory has unbroken supersymmetry, but (as we now describe) the relevant cosmological state breaks supersymmetry due to the presence of the time-varying scalar. For supersymmetric effective gravitational theories arising from string theory, it is also important to ensure that the scale of the compact extra dimensions is small. Examples of supersymmetric AdS string vacua with acceptably small cosmological constant and small extra dimensions have been described recently in \cite{Demirtas:2021nlu,Demirtas:2021ote}.

\paragraph{Symmetry breaking.}

The scalar field expectation values will typically break supersymmetry and some of the gauge symmetry. To understand this in detail, it is useful to rescale the scalar field in the action (\ref{scalaraction}) as $\phi = \sqrt{8 \pi G} \phi_p$ to give  standard particle physics normalizations. This gives
\be
\small{S_\phi = \int d^4 x \sqrt{g} \left[-\frac{1}{2} g^{ab}\partial_a \phi_p \partial_b \phi_p - \frac{\hat{m}^2}{L^2}  \phi_p^2 -\frac{1}{k L^2} V_{int}( \sqrt{k} \phi_p )\right]} \; .
\ee
Here, the quadratic term is set by the cosmological constant scale, and the dimensionless quartic coupling is naturally of order $k/L^2\sim G/L^2$. With this action, the potential varies over values of order the extremum of the potential (the negative cosmological constant associated with the AdS vacuum) for Planck scale variation of the field expectation values. Thus, we can have gauge symmetry and supersymmetry breaking at a high scale (set by the value of the scalar field) while the vacuum energy remains small. Fermions that couple directly to such a varying scalar via $\phi \psi \psi$ terms will have large time-dependent masses induced by the background value of $\phi$. The light fermions of the Standard Model will be some subset of the remaining fermions that do not get large masses in this way. The fluctuations $\delta \phi$ of the scalar about its time-dependent classical solution will correspond to a light scalar particle (whose mass is given by $V''(\phi)$), but we will only have simple $\delta \phi \psi \psi$ interactions for the fermions that also get a large mass from the background value of the scalar, so long range forces mediated by $\delta \phi$ on Standard Model matter might be avoided. Quanta of the $\delta \phi$ field will presumably contribute to the dark matter.

\paragraph{A new perspective on naturalness}

The framework provides a potential resolution for a number of naturalness problems in cosmology. Typically, these problems (e.g. the horizon and flatness problems, and the cosmological coincidence problem) are observations that the state of the universe at early times appears to be finely tuned in various ways in order that it can evolve to something that agrees with present observations. For example, this initial state must be extremely flat and homogeneous and have correlations between regions of space that (according to a simple evolution from the big bang) were never in causal contact. Inflation proposes a resolution by postulating an earlier phase of evolution that naturally leads to a state with these special properties. 

In our framework, the ``initial state'' naturally produced by the model is defined at the time-symmetric point where $\dot{a} = 0$. Thus, in addressing questions of naturalness, we should not ask if the state in the early universe is natural, but rather whether the state at the time-symmetric point (obtained by evolving forward from our present observed universe) is naturally produced by the Euclidean path integral in the model. The construction does give rise to a universe that is very flat and homogeneous because of the $\mathbb{R}^3$ symmetry of the model, and it leads to correlations on all scales, since the correlators are the same as vacuum correlators in the wormhole geometry in which there are massless fields. Thus, the framework may alleviate the need for inflation in resolving naturalness puzzles. However, it is important to understand whether the model can reproduce quantitative predictions of inflationary scenarios, namely the nearly scale-invariant spectrum of perturbations that gives rise to structure and CMB anisotropies that agree with observations.

Our framework also provides a possible explanation for the ``cosmic coincidence'' puzzle, that the vacuum energy and matter/radiation energy are of the same order of magnitude today despite evolving very differently under cosmological evolution (constant vs $1/a^3$ or $1/a^4$). In the flat recollapsing universes of our framework, the Friedmann equations imply that the total energy density must be exactly zero at the middle time when $\dot{a}=0$, so the negative contribution of dark energy must have exactly the same magnitude as the remaining positive contributions from matter/radiation. The time scale for the variation of the ratio of densities is the scale of dark energy at this time-symmetric point, which also sets the age of the universe, so the dark energy density and matter energy density will typically be of the same order of magnitude for most of the age of the universe. Provided that the present era is at a typical time, we will observe a coincidence in the energy densities.

\paragraph{Discussion.}

We have presented a possible framework for cosmology that should give specific predictions for the evolution of the background cosmology and its perturbations based only on the choice of four-dimensional gravitational effective theory (including a choice of boundary physics in the Euclidean picture). It remains to be seen whether realistic models of this type can be constructed, but in any case, the framework may provide a useful theoretical laboratory for learning about quantum aspects of cosmology.

\paragraph{Acknowledgements.}

We would like to thank Panos Betzios, Matt Kleban, Lampros Lamprou, Henry Maxfield, Yasunori Nomura, Liam McAllister, Douglas Scott, Eva Silverstein, Chris Waddell, David Wakeham, and Aron Wall for helpful discussions. This work is supported in part by the National Science and Engineering Research Council of Canada (NSERC) and in part by the Simons foundation via a Simons Investigator Award and the ``It From Qubit'' collaboration grant. PS is supported by an NSERC C-GSD award. SA is partially supported by a Leon A. Herreid Science Graduate Fellowship Award. This work is partially supported by the U.S. Department of Energy, Office of Science, Office of Advanced Scientific Computing Research, Accelerated Research for Quantum Computing program ``FAR-QC'' (SA) and by the AFOSR under grant number FA9550-19-1-0360 (BS).

\bibliographystyle{apsrev4-1}
\bibliography{references}

\end{document}